\documentclass{osa-article}

\journal{osac}
\usepackage{xcolor}

\articletype{Research Article}

\usepackage{lineno}
\usepackage{subcaption}

\begin{document}

\title{Scintillation Minimization versus Intensity Maximization in Optimal Beams}

\author{Anjali Nair,\authormark{1,*} Qin Li,\authormark{1}  and Samuel N Stechmann\authormark{1,2}}

\address{\authormark{1}Department of Mathematics, University of Wisconsin - Madison, Madison, WI 53706 USA\\
\authormark{2}Department of Atmospheric and Oceanic Sciences, University of Wisconsin - Madison, Madison, WI 53706 USA}

\email{\authormark{*}nair25@wisc.edu} 



\vspace{12pt}
\noindent
\begin{abstract}
In free-space optical communications and other applications,
it is desirable to design optical beams that have reduced or even
minimal scintillation.
However, the optimization problem for minimizing scintillation
is challenging, and few optimal solutions have
been found.
Here
we investigate the general optimization problem of minimizing scintillation and formulate it as a convex optimization problem. An analytical solution is found and demonstrates that a beam that minimizes scintillation is incoherent light (i.e., spatially uncorrelated). Furthermore, numerical solutions show that beams minimizing scintillation give very low intensity at the receiver. To counteract this effect, we study a new convex cost function that balances both scintillation and intensity. We show through numerical experiments that the minimizers of this cost function reduce scintillation while preserving a significantly higher level of intensity at the receiver.
\end{abstract}

\section{Introduction}

As optical beams propagate through the atmosphere, they undergo unwanted distortions. These distortions could be in the form of, for instance, intensity reduction, beam wander or scintillation at the receiver~\cite{andrews2005laser}. While fully coherent beams maximize the total average intensity at the receiver~\cite{schulz2005optimal}, partially coherent beams are known to have reduced scintillation. A lot of effort has been made in studying different types of beams with the aim of increasing intensity or reducing scintillation~\cite{andrews2005laser, schulz2005optimal, wu1991coherence, gbur2002spreading, dogariu2003propagation, feizulin1967broadening, korotkova2004model, baykal1983scintillations, baykal2009scintillations, gu2009scintillation, gu2010scintillation, garnier2022partially, polynkin2007optimized, wang2013experimental, avramov2014polarization, korotkova2014scintillation, Rabinovich:15,Rabinovich:16,Beason:20, li2022computation_conf, li2022optimal, li2022computation}.

Given the desire to reduce scintillation, it is also natural to try to minimize it.  The mathematical formulation for scintillation minimization as an optimization problem was proposed in~\cite{schulz2005optimal}, and it was shown theoretically that an optimal beam that minimizes scintillation is partially coherent. However, further details of the optimal, partially coherent beam are difficult to discern, since the optimization problem is challenging. For instance, it has been unclear as to how to compute the mutual intensity function $J$ that characterizes the optimal beams. 

Here, we consider the optimization problem for minimizing scintillation, and aim to overcome the difficulties of the optimization problem. 
In what follows, we first present the mathematical setup and
notation, and then demonstrate that scintillation-minimization can be formulated as an optimization
problem that is convex. Then we describe numerical methods
that are tractable and allow the optimal beam to be computed.
In earlier work,
optimal beams had been found when restricted to a special class of beams,
such as Gaussian Schell model beams
\cite{liu2006optimal,voelz2009metric,borah2010spatially,gbur2014partially,wang2015propagation}.
Here, moving beyond special classes of beams, we look for the
mutual intensity function $J$ among general beams.
For solving this general optimization problem, we present both numerical
and analytical results.

\section{Mathematical Setup and Notation}
The mathematical setup and notation are as follows.
Consider a source $\phi(X)$ located in a transmitter region $X\in\mathcal{A}$, and denote $I_0$ as the total average intensity at this transmitter. Let $U(X')$ be the field in the receiver region at $X'\in\mathcal{R}$. The received field $U$ is related to the source $\phi$ as 
\begin{equation}
    U(X')=\int\limits_{X\in\mathcal{A}}h(X,X')\phi(X)\mathrm{d}X\, ,
\end{equation}
where $h(X,X')$ is the propagator function.
Both the source and medium can have randomness, but they are assumed to be independent. The randomness of the medium is encoded in $h$. Then the total intensity at the receiver is
\begin{equation}\label{eqn:I}
    I=\int\limits_{X_{1},X_{2}\in\mathcal{A}}\langle\phi(X_1)\phi^\ast(X_2)\rangle H(X_1,X_2)\mathrm{d}X_1\mathrm{d}X_2 \, ,
\end{equation}
where
\begin{equation}\label{eqn:H}
    H(X_1,X_2)=\int\limits_{X'\in\mathcal{R}}h(X_1,X')h^\ast(X_2,X')\mathrm{d}X'\,
\end{equation}
is a Hermitian semi positive definite kernel, and $\langle\cdot\rangle$ is an expectation with respect to the randomness in the source. 

Two statistical quantities of interest are the mean and variance of this total intensity $I$. The mean total intensity can be computed by taking the average of~\eqref{eqn:I}:
\begin{equation}\label{eqn:E_I}
    \mathbb{E}[I]=\int\limits_{X_1,X_2\in\mathcal{A}}\mathbb{E}[H](X_1,X_2) J(X_1,X_2)\mathrm{d}X_1\mathrm{d}X_2\,,  
\end{equation}
with
\begin{equation}\label{eqn:J}
    J(X_1,X_2)=\langle\phi(X_1)\phi^\ast(X_2)\rangle\,.
\end{equation}
The quantity $J$ is a Hermitian semi positive definite function called the mutual intensity function~\cite{wolf1982new}. Here, $\mathbb{E}[\cdot]$ is an expectation with respect to the randomness in the medium. The scintillation index reflects the ratio of the variance and the intensity, namely:
\begin{equation}\label{eqn:S_old}
    \mathcal{S}=\frac{\mathrm{Var}[I]}{\mathbb{E}[I]^2}=\frac{\mathbb{E}[I^2]}{\mathbb{E}[I]^2}-1\,.
\end{equation}
The quantity $\mathbb{E}[I]/I_0$ characterizes the beam efficiency, and $\mathcal{S}$ characterizes the distortion of the beam due to the random medium (e.g., due to atmospheric turbulence).

\section{Optimization problem for scintillation minimization}

Next, in considering the optimization problem for minimizing scintillation,
a first question is whether or not it is convex.
Based on \eqref{eqn:S_old}, it is not clear whether 
scintillation is a convex function of $I$ or $J$ or $\phi$.
We now show that the optimization problem can be reformulated
and seen to be convex. Consequently, due to convexity,
the optimization problem is guaranteed to have a minimum,
and many computational algorithms that are designed for convex
problems can potentially be utilized.

To reformulate the minimization of \eqref{eqn:S_old} in manifestly convex form,
one can note a similarity to the Rayleigh quotient from
linear algebra: 
\textcolor{black}{the scintillation in \eqref{eqn:S_old} is unchanged if
$I$ or $J$ is rescaled by a constant factor. }
As a result,
following a brief calculation shown in the Supplementary Materials,
and substituting~\eqref{eqn:I} in~\eqref{eqn:S_old},
we can write the problem of scintillation minimization in discrete form as the following constrained convex optimization:
\begin{equation}\label{eqn:cost_old}
\begin{aligned}
        &\min_{J}  \quad J^\ast A J\quad \text{s.t. } \text{Tr}(\mathbb{E}[H] J^T)=1,\quad J\succcurlyeq 0
\end{aligned}
\end{equation}
where 
\begin{equation}\label{eqn:A}
    A=\mathbb{E}[H(X_1,X_2)H^\ast(X_3,X_4)]\,.
\end{equation}
Note that the notation in \eqref{eqn:cost_old} treats $J$ as a discrete quantity, so that $\text{Tr}(\mathbb{E}[H] J^T)$ is the discrete version of \eqref{eqn:E_I} as a Frobenius inner product, and $J^\ast AJ$ is the quadratic form
$\sum_{ijkl} J^\ast_{ij}A_{ijkl}J_{kl}$. This cost function $J^\ast AJ$ is convex with respect to the variable $J$. Since the constraint also forms a convex cone, the full problem can now be seen to be a convex minimization. Note that, while the reformulation and constraint have removed the non-uniqueness due to rescaling of $J$ by a constant factor, the minimizer may still be non-unique due to a different aspect: the operator $A$ is nonnegative definite but may not be strictly positive definite. 

For numerical calculations,
when the problem has a moderate size, one can also use CVX, a package for specifying and solving convex programs~\cite{cvx, gb08}, and in general cases, one can use the projected gradient descent algorithm~\cite{bottou2018optimization}, see also the iterative solver in~\cite{boyd2005least}.
For a 2-D simulation, $\mathcal{A}$ and $\mathcal{R}\subset\mathbb{R}$, and $A$ is a 4-dimensional tensor, and in 3-D a similar argument implies that $A$ is $8$-dimensional. This poses a serious requirement on the computer memory. On top of the memory issue, the computational requirement is also prohibitive. Since $A$ is the mean of $HH^\ast$, with the expectation taken on the random field, to compute $A$, one would need to run Monte Carlo simulation with many realizations of the random field, to make an ensemble average as an estimate to $A$. Moreover, each realization of $HH^\ast$ amounts to calling the propagator $h$ four times, each of which stores the full information of the associated Green's function. This is an infeasible numerical task for a brute-force computation.

To alleviate both computational and memory challenges, several computational strategies are employed here. The foremost technique is the incorporation of the randomized SVD solver~\cite{halko2011finding} that finds the eigenvalue/eigenfunction structure of $A\approx V\Sigma V^\ast$. There are two main advantages to utilize this solver. Firstly, the solver finds the eigenfunctions with a reduced cost. It has a quadratic dependence on the size of $A$ instead of cubic in conventional methods. A more attractive feature of this solver is that it does not require one to prepare $A$ ahead of time, but instead, only the knowledge of the action of $A\omega$, for a given vector $\omega$. In our particular case, this amounts to solving the field (corresponding to acting $H$ and $H^\ast$ on a vector) a couple times (corresponding to Monte Carlo sampling of the field) and take an average, completely removing the task of preparing the propagator. Further details of the numerical algorithms are described in the Supplementary Materials.

This strategy has enabled some reasonable computations. We simulate the field using the paraxial wave equation (PWE) with the frequency \textcolor{black}{$k=2\pi\times10^6$ rad/m and a propagation distance of $Z=3000$m.} A simple sinusoid function is used to represent the random potential $V(x,z)=V_1\sin(\omega_{x}x)\sin(\omega_{z}z)$. A splitting method~\cite{bao2002time} is adopted for simulating the PWE, and the standard cvx routine is called to find the optimal $J$. 


     \begin{figure}[htbp]
\begin{subfigure}{0.5\textwidth}   
    \centering
    \includegraphics[width=3.5cm]{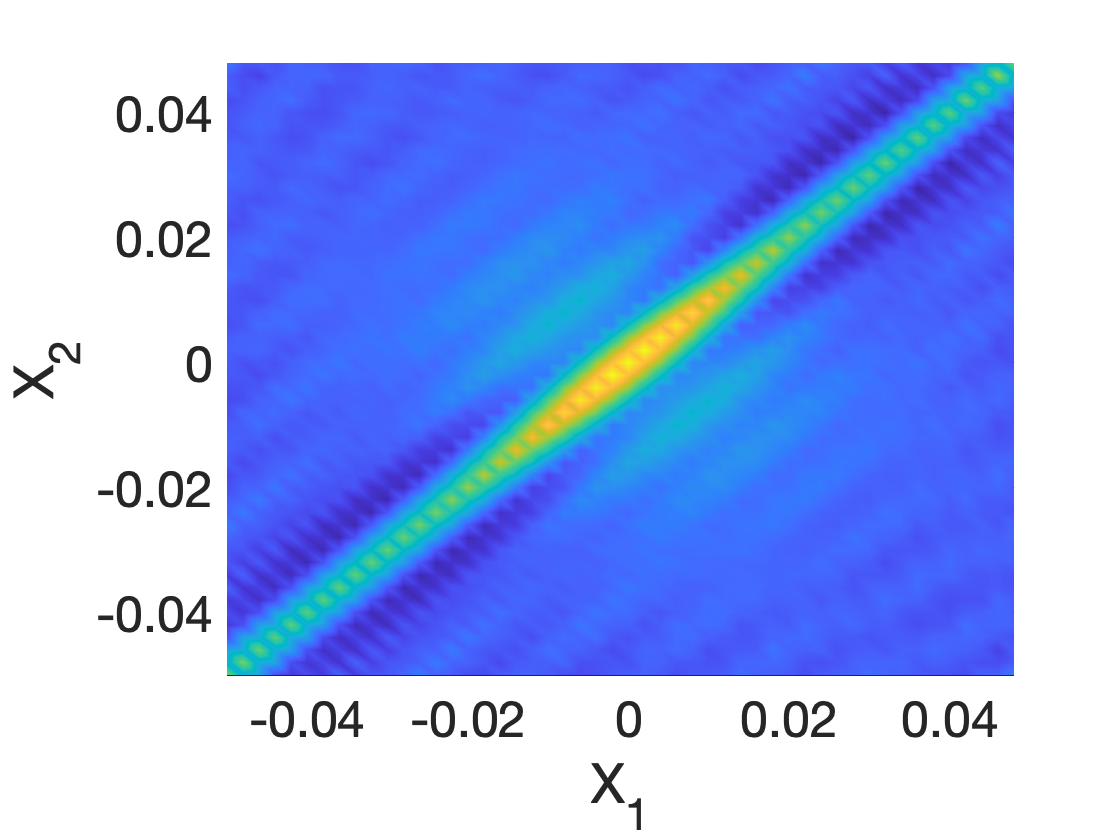}
    \caption{}
    \label{fig:J_optimal_PDE_real_mu=0}
\end{subfigure}
\begin{subfigure}{0.5\textwidth}    
    \centering
    \includegraphics[width=3.5cm]{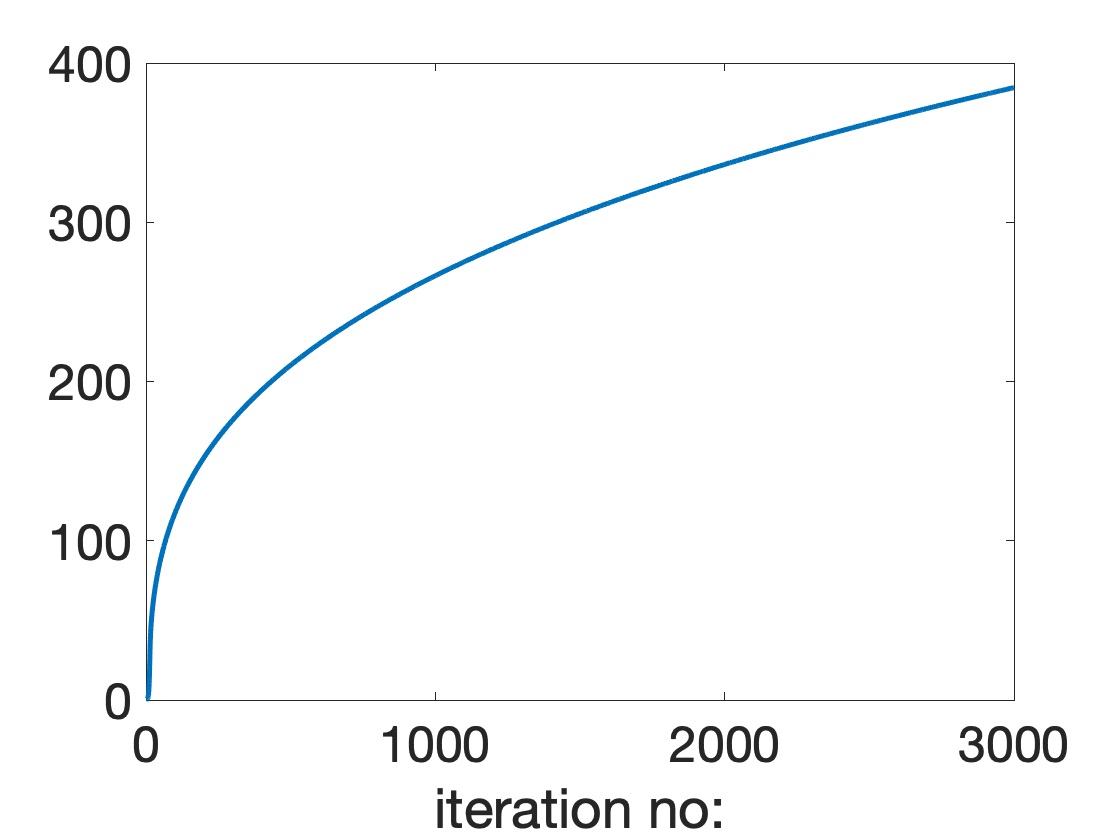}
    \caption{}
   \label{fig:eg_4_phase_screem_mu_0}
    \end{subfigure}
    \caption{Two illustrations that the scintillation-minimizing $J$ is a Dirac-delta function. (a) Profile of optimal $J$ in the non-parametrized case for PWE model. The optimal $J$ resembles a delta function. (b) Evolution of parameter $\lambda$ for a Gaussian parametrized $J=\exp\big(-\lambda^2(X_1-X_2)^2/2\big)$, and the \textcolor{black}{multiple} phase screen model. The value of $\lambda$ approaches infinity as the gradient descent progresses, meaning $J$ approaches a delta function.
    }
    \label{fig:J_optimal_mu=0}
\end{figure}


A plot of the optimal $J$ is shown in Fig~\ref{fig:J_optimal_PDE_real_mu=0}.
One interesting finding from Fig.~\ref{fig:J_optimal_PDE_real_mu=0} is that the optimal $J$ gives a delta-like function. This implies that the beam should be fully incoherent (white noise) to achieve the minimum scintillation. 

\textcolor{black}{As a second illustration of the same finding, 
we use a multiple phase screen model instead of the PWE.
The setup is a 2 dimensional problem with $k=2\pi\times10^6$rad/m, $ Z=2000$m, $r=0.04$m with 15 phase screens and a point receiver.}
Also, rather than allowing $J$ to be general, $J$ is parameterized of the form $\exp\big(-\lambda^2(X_1-X_2)^2/2\big)$.
Panel b (Fig.~\ref{fig:eg_4_phase_screem_mu_0}) shows the parameter $\lambda$ as gradient descent progresses. In this case, the optimal $J$ is found when $\lambda=\infty$, implying the fully incoherent beam is optimal in reducing scintillation.

The observation is in agreement with the physical intuition that partial coherence brings a reduction in scintillation, and here, taken to the extreme, it is seen that a minimum in scintillation is achieved by complete incoherence. 

\section{Analytical Result}

In addition to the numerical examples above,
we are also able to mathematically rigorously demonstrate this observation, in the setting of the random phase screen model of turbulence.
\textcolor{black}{In the case of a single phase screen model},
the propagator function takes the form
\begin{equation}\label{eqn:U_eHF}
h(X,X')=h_0(X,X')e^{i\psi(X)}\,,
\end{equation}
where $h_0(X,X')$ denotes the propagator for a uniform medium. Here $\psi$ is a random phase such that $\psi(X'_1)-\psi(X'_2)$ is a stationary random process with mean $0$ and covariance $D_\psi$~\cite{roggemann1996imaging}, the structure function of turbulence. One of the most commonly used structure functions is a power law of the form $D_\psi(X_1,X_2) = 6.88 (\|X_1-X_2\|/r_0)^{5/3}$, where $r_0$ is the Fried parameter of atmospheric turbulence. This formulation drastically simplifies the computation of $\mathbb{E}[H]$ and $A$~\cite{liu2006optimal}.

Now we would like to consider a source that is incoherent light, so that the mutual intensity $J$ is a Dirac delta function, and we now show by direct calculation that the scintillation is zero and therefore is minimized. To consider an approximate form of a Dirac delta function, let $\epsilon$ denote a small correlation length scale, and consider
\begin{equation}
J(X_1,X_2)=\frac{\mathbb{I}_{\epsilon}(X_1-X_2)}{H_0(0)\epsilon|\mathcal{A}|}, 
\label{eqn:J-phase-screen}
\end{equation}
where $\mathbb{I}_{\epsilon}$ is an indicator function that takes value $1$ if $|X_1-X_2|<\epsilon$ and a value of 0 otherwise, and $|\mathcal{A}|$ denotes the area of the transmitter region. \textcolor{black}{$H_0$ represents the value of $H$ in uniform medium and can be calculated explicitly~\cite{li2022computation}. }This $|\mathcal{A}|$ scaling leads to an initial intensity $I_0=\frac{1}{H_0(0)\epsilon}$. For small $\epsilon$, to leading order in $\epsilon$, a calculation shows
\begin{equation}\label{eqn:eHF_intensity}
\mathbb{E}[I]=1\,,\quad\text{and}\quad \mathbb{E}[I^2]=1\,,
\end{equation}
leading to a scintillation $\mathcal{S}=0$. See the Supplementary Materials for details of the calculation. This means that a completely incoherent beam provides an optimal beam when scintillation is the only criterion. 

Note that, for this optimal beam, the ratio $\mathbb{E}[I]/I_0$ of received intensity versus transmitted intensity is small and $O(\epsilon)$, so the ``beam'' of incoherent light will transmit a very low intensity signal that is practically useless. Also, note that this optimal $J$ in \eqref{eqn:J-phase-screen} can be rescaled by any constant factor (due to the form of the scintillation function in \eqref{eqn:S_old}) and will still lead to zero scintillation. Hence, if the transmitted intensity $I_0$ is rescaled to be $O(1)$ then the received intensity $\mathbb{E}[I]$ is $O(\epsilon)$.

{One can also show an elegant proof
(see Supplementary Materials for details) that the variance of $I$ is zero in the case when $J$ is a Dirac-delta function, under a quite general scenario for the turbulence model. However, the initial intensity $I_0$ in this case has infinite power and is not correctly mathematically defined (since it involves the evaluation of $\delta(0)$, the evaluation of a Dirac-delta function at the origin); and the finite-epsilon case in \eqref{eqn:J-phase-screen}--\eqref{eqn:eHF_intensity} clarifies the meaning of the (infinite) initial intensity $I_0$ in the $\epsilon\to 0$ limit.}

\textcolor{black}{If the one phase screen model is replaced by a multiple phase screen model, it can be shown that a Dirac-delta $J$ still achieves near zero scintillation, provided the transmitter is sufficiently large. See Supplementary Materials for details.}




Clearly from the examples above, reducing scintillation and amplifying intensity are goals pointing into opposite directions. The ``optimal'' beam in the sense of minimizing scintillation happens to be the worst beam in terms of preserving light intensity. 
Some previous work has also noted the importance of both intensity and scintillation in the cost function or metric, and has found optimal beams via numerical computation, for beams of a special class \cite{voelz2009metric,borah2010spatially,gbur2014partially,wang2015propagation}. Here we provide advances in the form of an analytical solution in \eqref{eqn:J-phase-screen}--\eqref{eqn:eHF_intensity}, and numerical solutions for beams of any general form of $J$.

These considerations motivate us to look for a modified cost function that balances scintillation and intensity efficiency. We use a modified objective function of
\begin{equation}\label{eqn:cost_new}
    \min_{J } \quad\mathcal{S}(J)+\mu\mathcal{Q}(J)\,,\quad\text{s.t.}\quad J\succcurlyeq 0\,.
\end{equation}
{where $\mathcal{Q}(J)=\big|\frac{I_0}{\mathbb{E}[I]}(J)-1\big|^2$ is a measure of the ratio of the transmitted and received intensities. A smaller value of $\mathcal{Q}$ suggests a smaller ratio $\frac{I_0}{\mathbb{E}[I]}$, meaning a larger amount of average received intensity.} 
As in~\eqref{eqn:cost_old}, any non-trivial constant scaling of the $J$ will leave the cost function invariant, so 
this alternative optimization problem can be reformulated
in a similar way to be manifestly convex.
This new cost function was chosen because it allows a
convex formulation, and because it balances the low scintillation at the receiver, and the high intensity efficiency, with $\mu$ being the balancing coefficient.



\section{Computational Results}

 \begin{figure}[htbp]
 
    \begin{subfigure}{0.5\textwidth}   
    \centering
    \includegraphics[width=4cm]{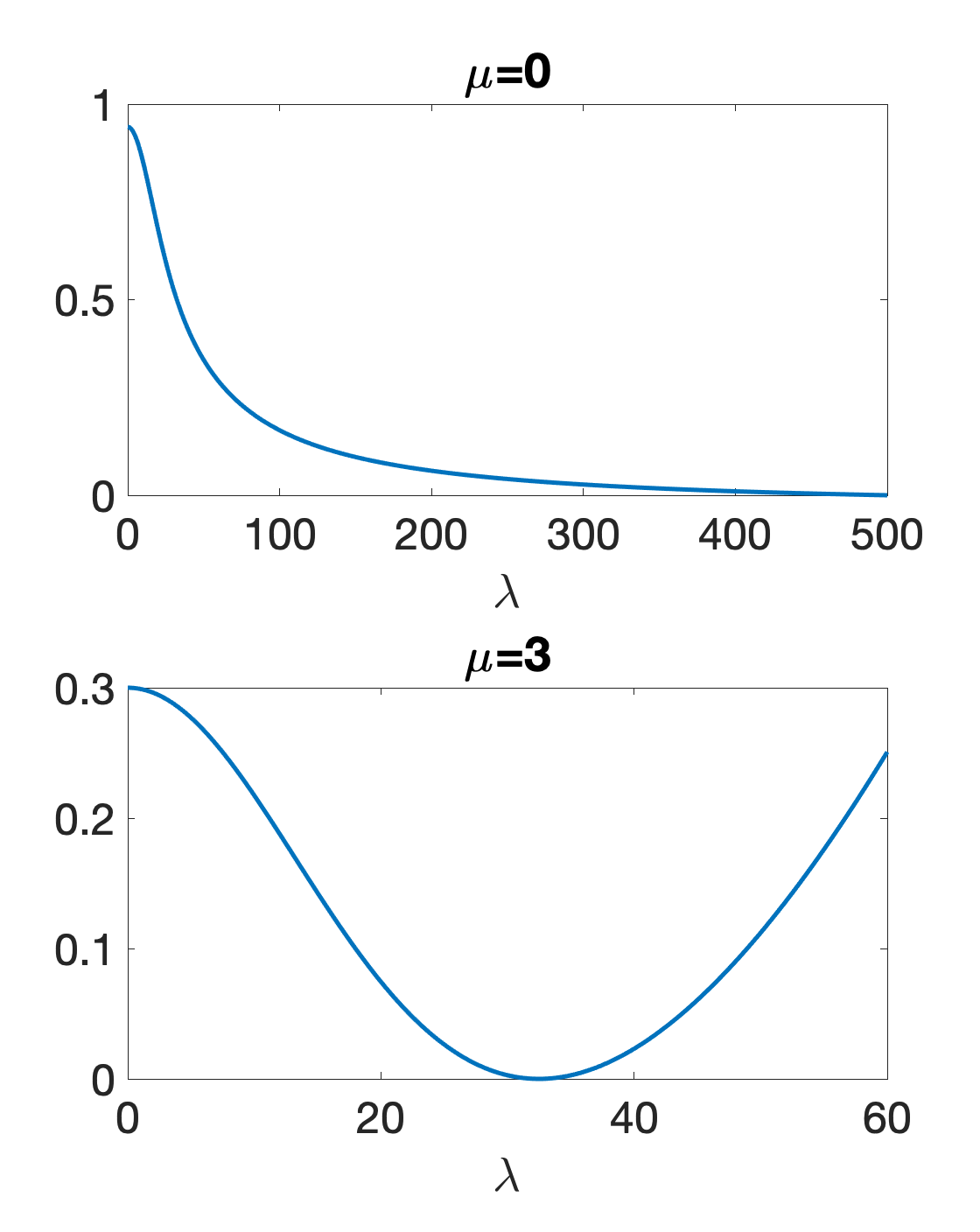}
    \caption{Cost function}
    \label{fig:eg4_phase_screen_cost_fn}
\end{subfigure}
\begin{subfigure}{0.5\textwidth}    
    \centering
    \includegraphics[width=4cm]{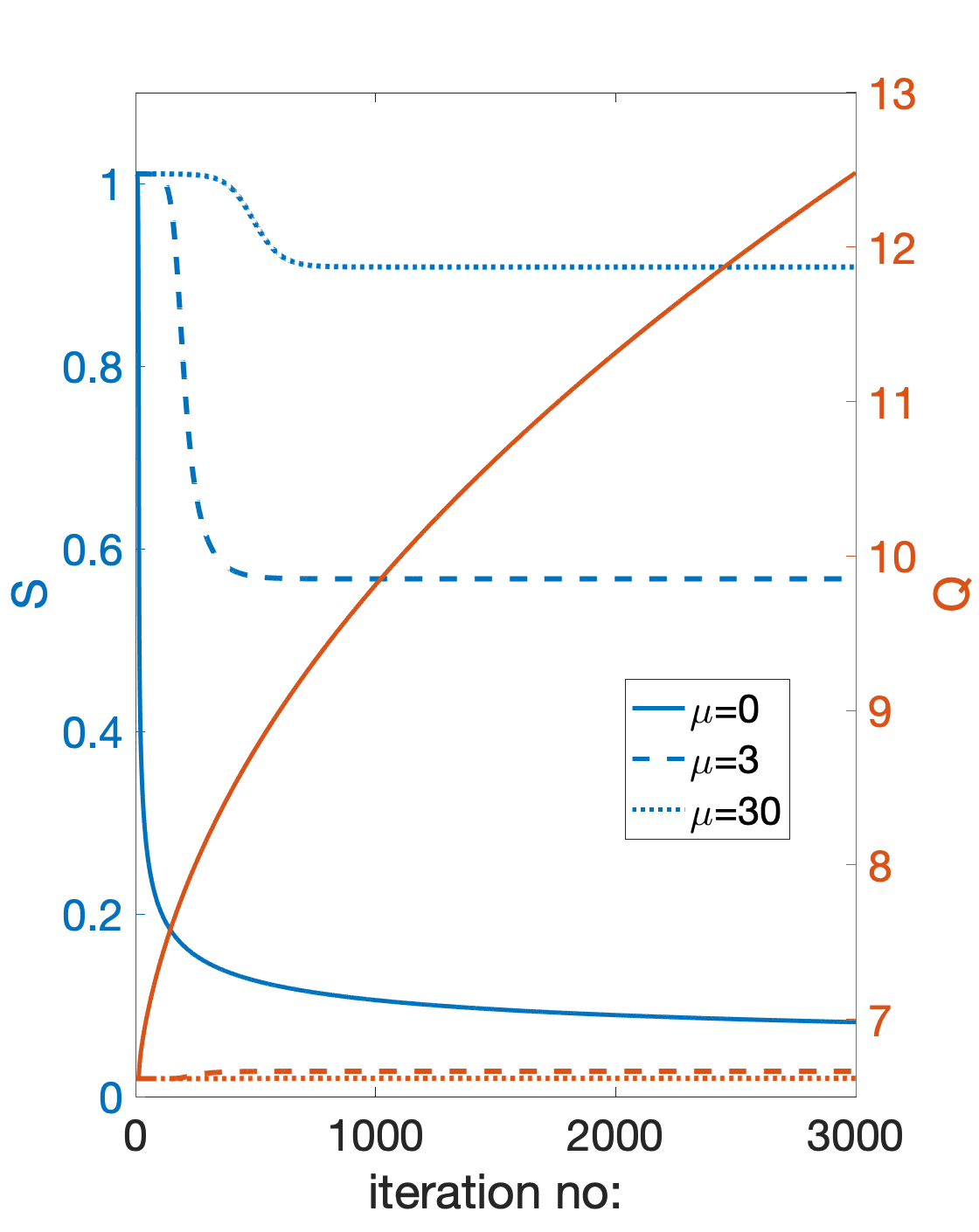}
    \caption{Scintillation and intensity}
   \label{fig:eg3_phase_screen_Gaussian}
    \end{subfigure}
    \caption{Alternative cost function from \eqref{eqn:cost_new} with $J$ parameterized as $\exp\big(-\lambda^2(X_1-X_2)^2/2\big)$. (a) Cost function in~\eqref{eqn:cost_new} as a function of $\lambda$, for different $\mu$. A unique minimum at finite $\lambda$ value is seen for $\mu\neq 0$. (b) Evolution of scintillation $\mathcal{S}$ and intensity quotient $\mathcal{Q}$ at each iteration of gradient descent. 
    Note that the $\mu=0$ case in Panel b has a larger stepsize.
    The iterations converge if $\mu\neq 0$.
    }
    \label{fig:Gaussian_J_GD}
\end{figure}

 \begin{figure}[ht]
\begin{subfigure}{0.5\textwidth}   
    \centering
    \includegraphics[width=3.5cm]{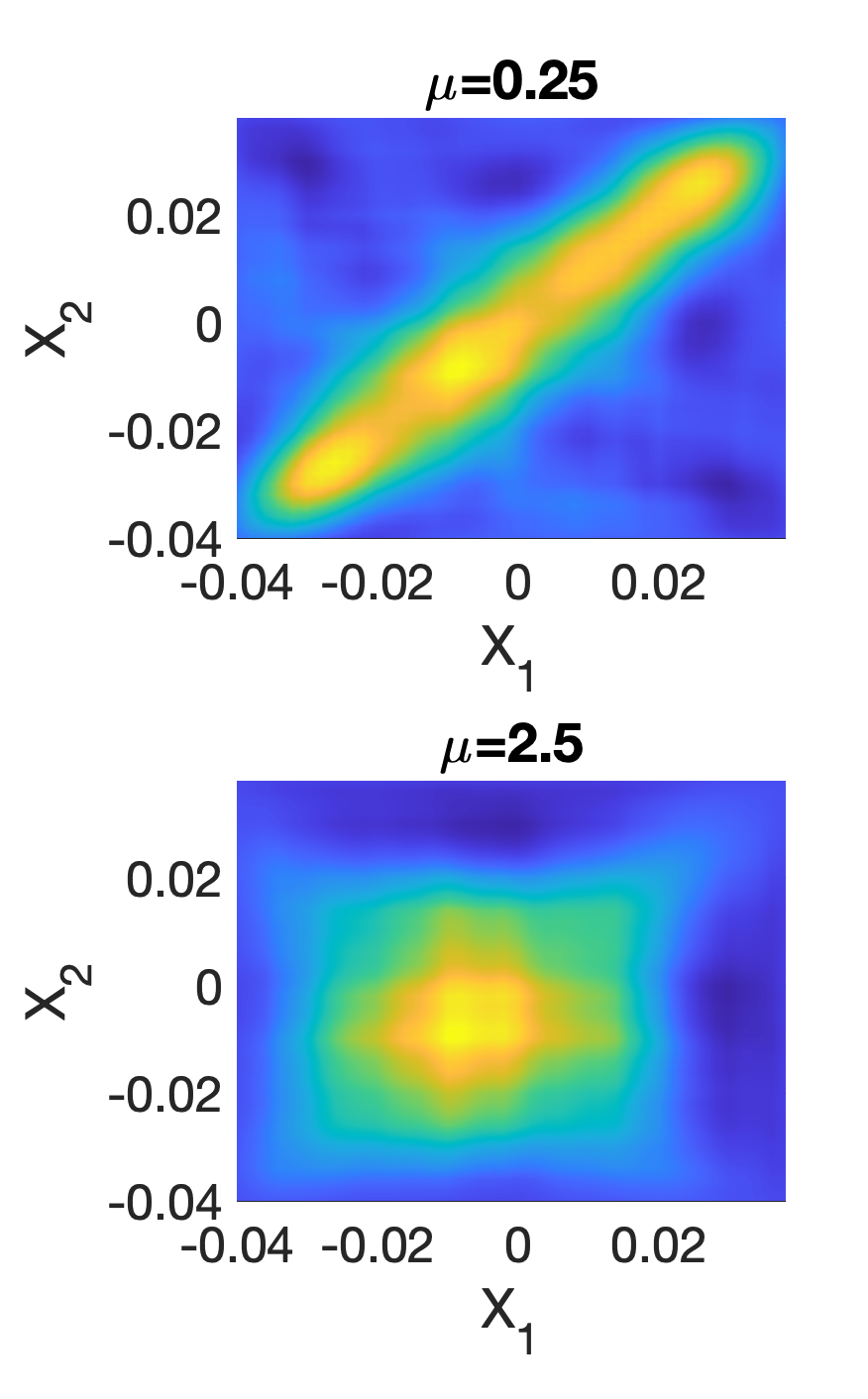}
    \caption{Phase screen model}
    \label{fig:J_phase_screen}
\end{subfigure}
 \begin{subfigure}{0.5\textwidth}    
    \centering
    \includegraphics[width=3.5cm]{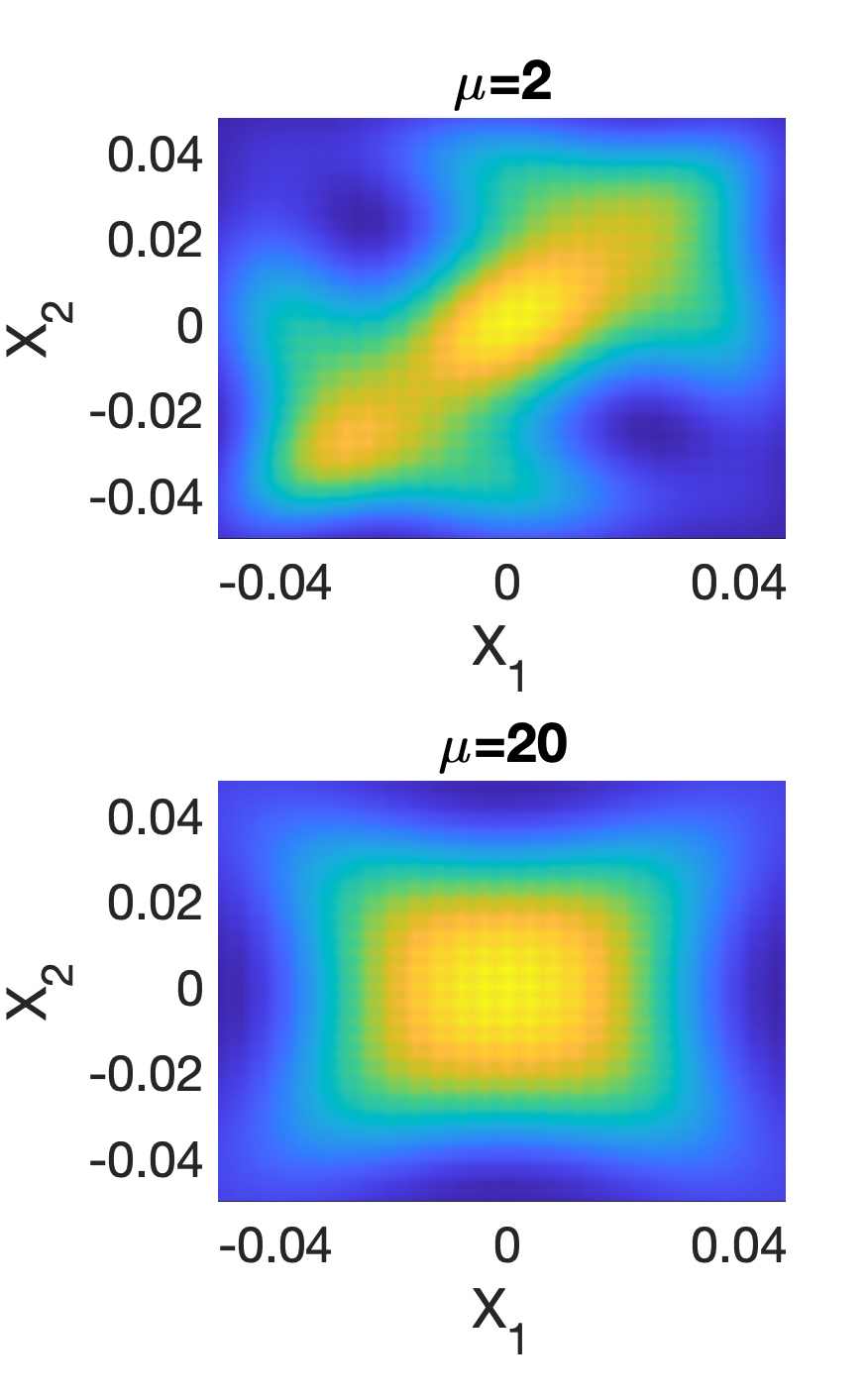}
    \caption{PDE model}
        \label{fig:J_PDE}
    \end{subfigure}  
    \caption{Profile of optimal $J$ (real part), where $J$ is allowed to be of general form and is non-parameterized. The left panel uses the phase screen model to compute the operator $A$ while the right panel uses PDE simulations. }
    \label{fig:optimal_J_mu_not_0}
\end{figure}

    \begin{figure}[ht]
\begin{subfigure}{0.5\textwidth}   
    \centering
    \includegraphics[width=3.5cm]{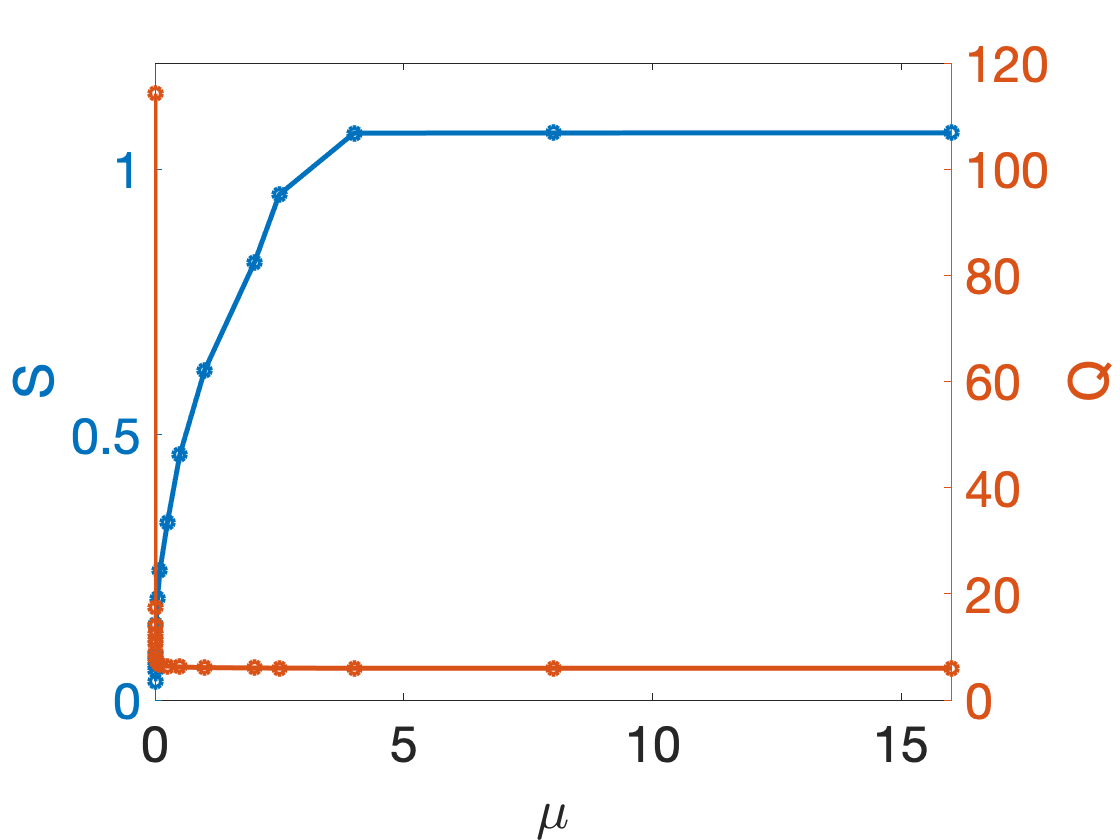}
    \caption{Phase screen model}
    \label{fig:scintillation_phase_screen}
\end{subfigure}
    \begin{subfigure}{0.5\textwidth}   
    \centering
    \includegraphics[width=3.5cm]{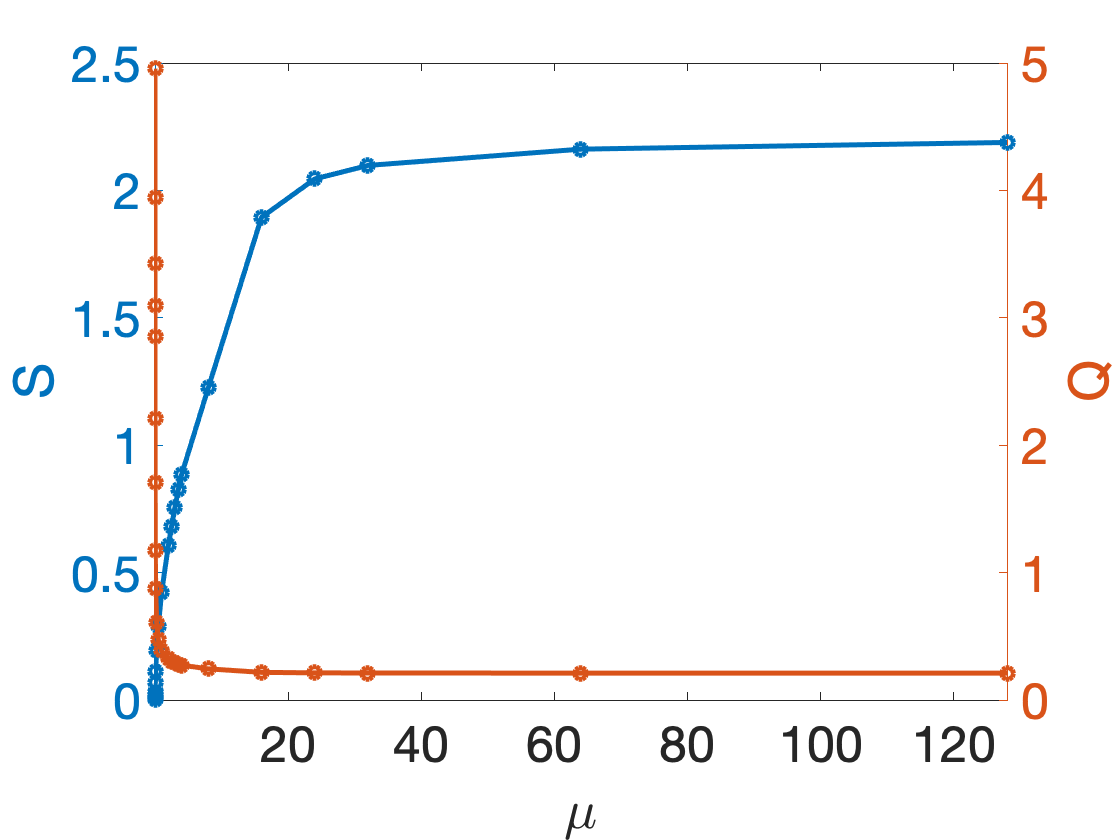}
    \caption{PDE model}
        \label{fig:scintillation_PDE}
\end{subfigure}
    \caption{Scintillation $\mathcal{S}$ and intensity quotient $\mathcal{Q}$ of optimal $J$.
    For small $\mu$ values (e.g., 
    \textcolor{black}{$\mu\approx 0.25$} in panel a and \textcolor{black}{$\mu\approx 2$} in panel b), the optimal $J$ provides substantial reduction in scintillation while maintaining adequate intensity. See the zoomed-in version in Supplementary Materials.
    }
    \label{fig:scintillation_and_intensity_general}
\end{figure}

We now evaluate the effect of such balancing via $\mu$ using numerical examples. To start, we use the phase screen model as set up for Fig.~\ref{fig:eg_4_phase_screem_mu_0} where $J$ is parameterized as  $\exp\{-\lambda^2(X_1-X_2)^2/2\}$. Figure~\ref{fig:eg4_phase_screen_cost_fn} shows the relationship between the objective~\eqref{eqn:cost_new} and $\lambda$ under different values of $\mu$. It is clearly seen that when $\mu=0$, the objective monotonically decreases and the minimizer is achieved for $\lambda=\infty$, whereas the minimizer is a finite $\lambda$ value whenever $\mu\neq 0$. Figure~\ref{fig:eg3_phase_screen_Gaussian} shows scintillation $\mathcal{S}$ and the quotient $\mathcal{Q}$ at different iterations in the direct use of Gradient Descent~\cite{boyd2005least}. A higher value of $\mu$ gives a larger scintillation and also a small quotient.

When $J$ is not parameterized but is allowed to take any general form, the optimal solution for different values of $\mu$ is plotted in Figure~\ref{fig:optimal_J_mu_not_0} using both the phase screen model and the PDE simulation. For both models, it is clear to see that a smaller value of $\mu$ favors a mutual intensity $J$ that is relatively diagonal, and partially coherent, whereas the optimal $J$ in the case of \textcolor{black}{a large $\mu$} is essentially coherent. The quantitative relation between the scintillation $\mathcal{S}$ and the quotient $\mathcal{Q}$ with respect to $\mu$ is plotted in Figure~\ref{fig:scintillation_and_intensity_general}. 
The value of $\mu$ controls a transition in the optimization problem from minimizing scintillation alone ($\mu=0$) to maximizing intensity with little consideration of scintillation \textcolor{black}{($\mu\gtrsim 2.5$ for the phase screen model and $\mu\gtrsim 20$ for the PDE model)}, and intermediate values of $\mu$ produce optimal beams with reduced scintillation and substantial intensity.
\section{Conclusion}

In conclusion, in this paper we investigated the optimization problem proposed in the literature \cite{schulz2005optimal} of finding the optimal beam that minimizes scintillation. We find that the scintillation-minimizing beam is incoherent light and has low intensity at the receiver. We find this result in both analytical solutions and numerical solutions, including cases where the mutual intensity function $J$ is non-parameterized and is allowed to be general. A modified objective function is introduced to balance the scintillation and the intensity. This optimization problem is convex. Utilizing machine learning algorithms (especially the randomized SVD solver) that exploit low-rank features, we can reduce both memory and computational cost and find the optimal mutual intensity function. This has been illustrated using numerical examples for Gaussian beams as well as general beams.

\section*{Funding} 
The research of Q.L. is partially supported by Office of Naval Research (ONR) grant N00014-21-1-2140, and the research of A.N. and S.N.S. is partially supported by ONR grant N00014-21-1-2119.

\section*{Acknowledgments} 
The authors thank Svetlana Avramov-Zamurovic and two anonymous reviewers for helpful comments and discussion. 

\section*{Disclosures} 
The authors declare no conflicts of interest.

\section*{Data Availability Statement} 
Data and code underlying the results presented in this paper are available in~\cite{Nair2022_SVD_code}.

\section*{Supplemental document}
Supporting content is available in \href{https://doi.org/10.6084/m9.figshare.21565545}{Supplement 1}.

\bibliography{Reference}

\end{document}